\input harvmac
\input epsf
\Title{ \vbox{\baselineskip12pt \hbox{HUTP-99/A066} \hbox{hep-th/9912002} } }
{\vbox{
\centerline{On the Uniqueness of Black Hole Attractors}
}}
\centerline{ Martijn Wijnholt and Slava Zhukov}
\vskip .1in
\centerline{\it Jefferson Laboratory of Physics, Harvard
University}
\centerline{\it Cambridge, MA 02138, USA}
\vskip .5in


\centerline{\bf Abstract}

\noindent 
We examine the attractor mechanism for extremal black holes in the context
of five dimen\-sional $N = 2$ supergravity and show that attractor points are
unique in the extended vector multiplet moduli space. Implications for
black hole entropy are discussed.

\vskip .3in
\Date{December 1999}

\newsec{Introduction}

BPS black holes in four and five dimensional $N = 2$ supergravity have been
much studied using the attractor mechanism. To construct a black hole
solution, one specifies the charges and asymptotic values of the
moduli. The moduli then evolve as a function of the radius until they reach
a minimum of the central charge at the horizon. This minimum value
determines the entropy of the black hole. Consequently, it is important to
know if different values of the central charge can be attained at different
local minima, or perhaps even more basically if multiple local minima are
allowed. If uniqueness fails, one would be led to believe that the
degeneracy of BPS states does not solely depend on the charges.

As it turns out, in the five dimensional case we can show that at most one
critical point of the central charge can occur; that is the subject of this
paper. The structure of the argument is simple. Remarkably, the extended
K\"ahler cone turns out to be convex. Therefore we can take a straight line
between two supposed minima and analyse the (correctly normalised) central
charge along this line. The central charge cannot have two minima when
restricted to this line, yielding a contradiction.

We would like to emphasize that although Calabi-Yau spaces will be in the
back of our minds for most of this paper, the geometric statements have
clear analogues in five dimensional supergravity, so that our arguments are
independent of the presence of a Calabi-Yau. We will point out some of the
parallel interpretations where they occur.

In section two we provide two arguments for uniqueness in a single K\"ahler
cone. In section three we tackle the extended K\"ahler cone, and in section
four we discuss some of the implications. The reader may wish to start with
section four before moving on to the arguments of sections two and three.

\newsec{Single K\"ahler cone}

\subsec{Review of the attractor mechanism}

Let us recall the basic setting for the five-dimensional attractor
problem. We consider M-theory compactified on a Calabi-Yau threefold. As
the low-energy effective theory we obtain five dimensional $N=2$
supergravity with $h^{1,1}-1$ vector multiplets, $h^{2,1}+1$
hypermultiplets and the gravity multiplet\thinspace
\ref\CCAF{ A.C.~Cadavid, A.~Ceresole, R.~D'Auria and S.~Ferrara,
``Eleven-dimensional supergravity compactified on Calabi-Yau threefolds,''
Phys.\ Lett.\ {\bf B357}, 76 (1995) hep-th/9506144.}.
The vector multiplets each contain one real scalar, so the vector moduli
space is $h^{1,1} - 1$ dimensional.  From the point of view of Calabi-Yau
compactification these scalars have a simple geometrical
interpretation. Let us denote the Calabi-Yau three-fold by $X$ and expand
two-cycles $Q$ of $X$ as $Q = Q_i \, e^i$, where $e^i$ is a basis for
$H_2(X,\bf{Z})$. The dual basis for $H^2(X,\bf{Z})$ will be written with
lower indices, $e_j$, so that $\left< e_j e^i \right> =\delta^i_j$.  The
Calabi-Yau K\"ahler class $k$ can be expanded as $k = k^i \, e_i$ which
gives an $h^{1,1}$ dimensional space of parameters. One parameter
corresponding to the total volume of the Calabi-Yau is part of the
universal hypermultiplet and the rest of the parameters corresponding to
the sizes of cycles in the Calabi-Yau describe precisely the moduli space
of vector multiplets.

The dynamics of vector multiplets in $N=2$ supergravity is completely
governed by the prepotential $F(k)$, which is a homogeneous cubic
polynomial in vector moduli co\"ordinates $k^i$.  It is a special property
of five dimensional supergravity that there are no nonperturbative quantum
corrections to this prepotential\thinspace
\ref\BBS{ K.~Becker, M.~Becker and A.~Strominger, ``Five-branes,
membranes and nonperturbative string theory,'' Nucl.\ Phys.\ {\bf B456}
(1995) 130 hep-th/9507158.}.
Geometrically the prepotential is simply the volume of $X$ in terms of~$k$
\eqn\prep{ F(k) \equiv {1\over 6}\int_X k \wedge k \wedge k = 
{1\over 6} k^i k^j k^l d_{ijl} }
where $d_{ijl}$ denote the triple intersection numbers of $X$
$$ d_{ijl} = \int_X e_i \wedge e_j \wedge e_l. $$
In order to abbreviate the formulae, let us introduce the
following notation:
$$\eqalign{ a \cdot b &\equiv a_m b^m  \cr
 a \cdot b \cdot c &\equiv a^i b^j c^l d_{ijl} \cr
 k^3 &\equiv k \cdot k \cdot k } $$
To decouple the universal hypermultiplet co\"ordinate we need to impose the
constraint $F(k)=1$ which gives us the vector moduli space as a
hyper\-surface inside the K\"ahler cone. Alternatively, we will sometimes
think of the moduli space as a real projectivisation of the K\"ahler
cone. In this case we have to consider functions invariant under overall
rescaling of $k$'s.

The prepotential defines the metric on moduli space as well as the gauge
coupling matrix for the five-dimensional gauge fields. Including the
graviphoton, there are exactly $h^{1,1}$ $U(1)$ gauge fields in the theory
and their moduli-dependent gauge coupling matrix is given by\thinspace
\ref\sugra{ M.~Gunaydin, G.~Sierra and P.K.~Townsend,
``The Geometry Of N=2 Maxwell-Einstein Supergravity And Jordan Algebras,''
Nucl.\ Phys.\ {\bf B242}, 244 (1984).}
\eqn\gauge{ G_{ij} = -{1 \over 2} { \del^2 \over \del k^i \, \del k^j }
         \log F(k) = -{1 \over 2} { \del^2 \over \del k^i \, \del k^j }
         \log k^3}
The moduli space metric $g_{ij}$ is just the restriction of $G_{ij}$ to the
hyper\-surface $F(k)=1$. In the supergravity Lagrangian $G_{ij}$ and
$g_{ij}$ multiply kinetic terms for the gauge fields and the moduli fields
respectively. It is then very important that both metrics should be
positive-definite inside the physical moduli space. The tangent space to
the $F(k)=1$ hyper\-surface is given by vectors $\Delta k$ such that $k
\cdot k \cdot \Delta k =0 $. Then positivity of $g_{ij}$ requires
\eqn\positive{ \Delta k^i \, g_{ij} \, \Delta k^j \equiv
               \Delta k^i \, G_{ij} \, \Delta k^j =
- 3 \, k \cdot \Delta k \cdot \Delta k  > 0 .}
\nref\SUSYandAttr{S.~Ferrara and R.~Kallosh, ``Supersymmetry and
Attractors,'' Phys.\ Rev.\ {\bf D54}, 1514 (1996) hep-th/9602136.}
\nref\FiveDims{A.~Chou, R.~Kallosh, J.~Rahmfeld, S.~Rey, M.~Shmakova
 and W.K.~Wong, ``Critical points and phase transitions in 5d
compactifications of M-theory,'' Nucl.\ Phys.\ {\bf B508}, 147 (1997)
hep-th/9704142.}

Next we consider BPS states with given electric charges.  The vector of
electric charges with respect to the $h^{1,1}$ $U(1)$ gauge fields can be
thought of as an element $Q$ of $H_2(X,\bf{Z})$. In M-theory language these
BPS states are M2-branes wrapped on a holomorphic cycle in the class
$Q$. For large charges we can represent them in supergravity by certain
extremal black hole solutions\thinspace
\ref\Sabra{W.A.~Sabra, ``General BPS black holes in five dimensions,''
Mod.\ Phys.\ Lett.\ {\bf A13}, 239 (1998) hep-th/9708103.}.\thinspace
The structure of these solutions is as follows.  As one moves radially
towards the black hole the vector multiplet moduli fields $k^i$ vary. They
follow the gradient flow of the function $Z \equiv Q \cdot k$:
$$\eqalign{
 \partial_{\tau} U &= + {1\over 6} e^{-2 U} Z  \cr
 \partial_{\tau} k^i &= -{1\over 2} e^{-2U} G^{ij} D_j Z }
$$
Here $U \equiv U(r)$ is the function determining the five dimensional
metric
$$ ds^2 = - e^{-4U} dt^2 + e^{2U} (dr^2 + r^2 d\Omega^2), $$
$\tau = 1/ r^2$ and the covariant derivative is
$$ D_j = \partial_j - {1\over 6} k^i k^l d_{ijl}. $$
Geometrically $Z$ is just the volume of the holomorphic cycle $Q$ in the
Calabi-Yau with K\"ahler class $k$. We will refer to $Z(k)$ as the central
charge because when evaluated at infinity, $Z$ is indeed the electric
central charge of the $N = 2$ algebra. As we approach the horizon of the
black hole at $r = 0$ or $\tau = \infty$, the central charge rolls into a
local minimum and the moduli stabilise there, let us call that point
$k_0$. The area of the horizon and thus the entropy of the black hole are
determined only by the minimal value of the central charge
\refs{\SUSYandAttr,\FiveDims} :
\eqn\entropy{ S = {\pi^2 \over 12} \, Z_0^{3/2} \, =
       {\pi^2 \over 12} \left( \int_Q k_0 \right)^{3/2} =
       {\pi^2 \over 12} (Q \cdot k_0)^{3/2} .}
Those points in the moduli space where the central charge attains a local
minimum for a fixed electric charge $Q$ are called attractor points.

The microscopic count of the number of BPS states with given charge has
been performed for the special case of compactifications of M-theory on
elliptic Calabi-Yau threefolds\thinspace
\ref\Vafa{ C.~Vafa, ``Black holes and Calabi-Yau threefolds,''
Adv.\ Theor.\ Math.\ Phys.\ {\bf 2}, 207 (1998)
hep-th/9711067.}.\
There the attractor point for any charge vector $Q$ was found explicitly
and the resulting entropy prediction \entropy\ agreed with the microscopic
count for large charges.

It was pointed out in\thinspace
\ref\Moore{G.~Moore, ``Arithmetic and attractors,'' hep-th/9807087.}
that in the case of general Calabi-Yau compactifications an attractor point
is not necessarily unique, and in principle for making an entropy
prediction one needs to specify not only the charges of the black hole, but
also an attractor basin, that is one needs to specify a region in moduli
space in which all the points flow to a given attractor point along a path
from infinity to the horizon. In the remainder of this article we show that
if a minimum of the central charge exists, the attractor basin for this
minimum covers the entire moduli space. There cannot be a second local
minimum and so the specification of the charges of the black hole is
sufficient for determining the attractor point.

\subsec{Geometric argument}

To find an attractor point explicitly, one needs to extremise the central
charge subject to the constraint ${k \cdot k \cdot k = 1}$, which leads
directly to the five dimensional attractor equation \FiveDims
$$ Q_i = (Q \cdot k)\, k^j\, k^l\, d_{ijl} $$
In differential form notation, it reads
\eqn\attr{ [Q] = \left( \int_Q k \right) [k \wedge k]. }
Here $[Q]$ is a four-form which is Poincar\'e dual to the two-cycle $Q$.
For convenience, we will leave out the square brackets in what follows.

Let us recall some standard facts about the Lefschetz decomposition (see
for instance\thinspace
\ref\GriffithsHarris{P.~Griffiths and J.~Harris, ``Principles of Algebraic
Geometry,'' Wiley-Interscience (1978).}). 
On any K\"ahler manifold the K\"ahler class is a harmonic form of type
(1,1). It can therefore be used to define an action on the cohomology. We
define the raising operator to be the map from $H^{p,q}(X, {\bf C})$ to
$H^{p+1,q+1}(X, {\bf C})$ obtained by wedging with $k$
$$ L_k \alpha = k \wedge \alpha $$
and similarly the lowering operator to be the map from $H^{p,q}(X, {\bf
C})$ to $H^{p-1,q-1}(X, {\bf C})$ obtained by contracting with $k$
$$ \Lambda_k \alpha = \iota_k \alpha.$$
The commutator sends forms of type (p,q) to themselves up to an
overall factor:
\eqn\su{ [L, \Lambda] = (p + q - n) {\bf I }}
where $n$ is the complex dimension of $X$. Thus $L$, $\Lambda$ and $(p + q
- n) {\bf I} $ form an $sl(2,{\bf R})$ algebra and the cohomology of $X$
decomposes as a direct sum of irreducible representations.  When $X$ is a
Calabi-Yau threefold, the decomposition is
$$ H^*(X,{\bf C}) = 1({\bf {3/2}}) \oplus 
    (h^{1,1} - 1)({\bf {1/2}}) \oplus (2 h^{2,1} + 2)({\bf 0}). $$ 
The spin 3/2 represenatation corresponds to $\{1, k, k^2, k^3\}$.  There
can be no spin 0 representations in $H^{1,1}(X,{\bf C})$ because if
$\alpha$ is of type (1,1) and $L_k \alpha = 0$ then by equation
\su\ we deduce that $\alpha$ is zero. In particular, the raising
operator $L_k$ maps classes of type (1,1) isomorphically onto classes of
type (2,2). We will use this fact in the following argument.\foot{ 
We thank C.~Vafa for this argument.}

To prove that a single K\"{a}hler cone supports at most one attractor
point, assume to the contrary that there are two such points, $k_0$ and
$k_1$, satisfying \attr. Then we may rescale $k_0$ or $k_1$ by a positive
factor such that
\eqn\sign{ k_0 \wedge k_0 = \pm k_1 \wedge k_1.}
First we fix the sign in the above equation. Since ${1\over 2}(k^0 + k^1)$
is inside the K\"ahler cone, $k^0 + k^1$ is an admissible K\"ahler class
and $\int_X (k^0 + k^1)^{\wedge 3}$ is positive. Assuming the sign in
\sign\ is minus, one may expand $(k_0 + k_1)^{\wedge 3}$ and deduce that
$$\int (k_0 + k_1)^{\wedge 3} =
-2\int (k_0)^{\wedge 3} -2\int (k_1)^{\wedge 3} < 0 \;,$$
which is impossible, so the sign is a plus. Therefore we have
$$(k_0 + k_1) \wedge (k_0 - k_1) = 0.$$
As discussed above, $L_{k_0 + k_1}$ cannot annihilate any classes of type
(1,1) because $k^0 + k^1$ is an allowed K\"ahler class.  We conclude that
$k_0 - k_1$ must vanish.

\subsec{Physical argument}

One doesn't really need the attractor equation to prove that in a single
cone multiple critical points cannot occur.  Another argument makes use of
the simple properties of the prepotential \prep\ and gives further insight
into the behaviour of the central charge function.

Let us examine the behaviour of the central charge along straight lines in
the K\"ahler cone.  First take any two points $k_0$ and $k_1$ in the
K\"ahler cone.  By convexity of the cone we can take a straight line from
$k_0$ to $k_1$,
\eqn\straight{ k(t) = k_0 + t \, \Delta k, \quad \Delta k =  k_1 -
k_0.}
For $t$ between $0$ and $1$ and a little bit beyond those values $k(t)$
certainly lies in the K\"ahler cone, but it no longer satisfies
$k(t)^3=1$. To cure this we think of the moduli space as a real
projectivisation of the K\"ahler cone and define the central charge
everywhere in the cone by normalising $k$:
\eqn\Zk{ Z(k) = { \int_Q k \over (\int_X k \wedge k \wedge k)^{1/3 }} =
                { Q \cdot k \over (k^3)^{1/3}} }
Then the central charge along the straight line \straight\ is just
\eqn\Zt{ Z(t)  = {Q \cdot k(t) \over (k(t)^{3})^{1/3}}. }
Let us also assume that the central charge is positive at $k_0$, i.e.
$Z(0)>0$. Otherwise we would consider the same problem with the opposite
charge.

Differentiating $Z(t)$, one finds for the first derivative
\eqn\firstder{ Z'(t) = (k^3)^{-4/3}
\left( (Q \cdot \Delta k)(k^3)
                  - (Q \cdot k)(\Delta k \cdot k \cdot k)\right).}
Now suppose $Z(t)$ has a critical point $t_c$ where $Z'(t_c) = 0$. Then the
second derivative at $t_c$ can be expressed as
$$\eqalign{
Z''(t_c) &= 2 (k^3)^{-4/3} (Q \cdot k)
            \left( {(\Delta k \cdot k \cdot k)^2 \over k^3}
                  - \Delta k \cdot \Delta k \cdot k \right) \cr
         &= 2 (k^3)^{-4/3} (Q \cdot k) (B_{ij} \, \Delta k^i
                                                   \Delta k^j).}$$
The bilinear form $B_{ij}$ has the following properties: exactly one of its
eigenvalues is zero (namely in the $k$-direction) and the other eigenvalues
are positive. In the language of Calabi-Yau geometry this holds\foot{See
for instance \GriffithsHarris, page 123. Note the misprint there; it
should say $2k = p + q$.}
because $\int_X k \wedge \Delta k \wedge \Delta k < 0$ for any $\Delta k$ that
satisfies ${\int_X k \wedge k \wedge \Delta k = 0}$.
Now if $\Delta k$ would be proportional to $k(t) = k_0 + t \Delta k$ for
some $t$, we would find that $k_0$ and $k_1$ are in fact equal. Thus
$\Delta k$ necessarily has a piece that is orthogonal to $k(t)$ and so
$$ B_{ij} \, \Delta k^i \Delta k^j > 0 \; \; {\rm strictly}.$$
In the language of supergravity the above statement follows from the fact
that the form $B$ is proportional to the metric when restricted to
directions tangent to the moduli space, for which $k \cdot k \cdot \Delta k
= 0$, see \positive. The zero eigenvalue in the $k$-direction is simply the
scale invariance of $Z(k)$.

The above inequality can be expressed in the following words: for positive
central charge {\it any critical point along any straight line is in fact a
local minimum!} For negative $Z$ every critical point along a straight line
is a local maximum. It is well known that a critical point of the central
charge is a minimum when considered as a function of all moduli. Here we
have a much stronger statement. We see that on a one-dimensional subspace
(a projection of a straight line) any critical point is in fact a local
minimum.

We can use the above observation as follows. The central charge $Z$ has a
local minimum at the attractor point $k_0$ by definition. Therefore it must
grow continuously on any straight line emanating from $k_0$ and can never
achieve a second local minimum. Moreover, we see that the central charge
has a global minimum at $k_0$ in the entire K\"ahler cone.

Let us remark on another consequence of our observation. Consider level
sets of the central charge function, i.e sets where $Z<a$ for some constant
$a$. Note that all such sets are necessarily convex. For otherwise, if we
could connect two points inside a level set by a line segment venturing
outside it, there would be a maximum of the central charge on that line
segment, which contradicts the above observation.

\newsec{Extended K\"ahler cone}

\subsec{Review}

The single K\"ahler cone we have just discussed is only a part of the full
vector moduli space.\negthinspace
\foot{See
\ref\Phases{E.~Witten, ``Phase Transitions In M-Theory And F-Theory,''
Nucl.\ Phys.\ {\bf B471}, 195 (1996) hep-th/9603150.}
for discussion. }
Some of the boundaries of the K\"ahler cone correspond to actual boundaries
of the moduli space. At other boundaries the Calabi-Yau undergoes a flop
transition, that is a curve collapses to zero size, but one may continue
through the wall and arrive in a different geometric phase. There one has
another Calabi-Yau which is birationally equivalent to the original
one. They share the same Hodge numbers but have different triple
intersection numbers. In terms of the original K\"ahler moduli, the
collapsed curve has a finite but negative area on the other side of the
wall. In five dimensions this has the interpretation of a phase transition
where a BPS hypermultiplet goes from positive to negative mass. The union
of the K\"ahler cones of all Calabi-Yaus related to each other through a
sequence of flop transitions is called the extended K\"ahler
cone. Geometrically one cannot go beyond the boundaries of this extended
cone. It has also been argued that at the boundaries that are at finite
distance the physical vector moduli space ends \Phases.

After we cross the wall into an adjacent cone we may take linear
combinations of K\"ahler parameters $k^i$ in order to get an acceptable set
of moduli that yield positive areas for two- and four-cycles. But we will
find it more convenient to stick to the original $k^i$, even though they
can sometimes yield negative areas outside the original cone. By induction,
we may still use the $k^i$ if we pass through a second flop transition into
a third cone, and so on.

The Calabi-Yau in the adjacent cone has different intersection numbers,
which means that we have to adjust the prepotential for the new cone.  It
is well known how the prepotential changes when one passes through a wall:
if we denote by $m$ the area of the collapsing curve that is negative on
the other side of the wall, then\thinspace
\ref\Morrison{D.~Morrison, ``Beyond the K\"ahler cone,'' Proc. of
the Hirzebruch 65 conference on algebraic geometry (M.~Teicher, ed.),
Israel Math. Conf. Proc., vol.~9, 1996, pp. 361--376, alg-geom/9407007.}
\eqn\change{ k\cdot k \cdot k \to k\cdot k \cdot k -  (\# {\bf P}^1)\ m^3 .}
Here $\# {\bf P}^1$ stands for the number of ${\bf P}^1$'s shrinking to
zero size at the wall.  Intuitively, the growing curve should contribute a
positive number to the volume of the Calabi-Yau for $m < 0$, hence the
minus sign in \change. This sign is crucial for proving uniqueness of
attractor points in the extended K\"ahler cone.

Physical quantities experience only a mild change at the flop transition
\FiveDims. In parti\-cular from \change\ we see that the prepotential is
twice continuously differentiable. The central charge $Z$ is also twice
continuously differentiable. The metric, which involves second
derivatives of the prepotential, is only
continuous.

\subsec{Convexity of the extended cone}

The extended K\"ahler cone has an alternative description in terms another
cone, as we will explain below. The advantage of this description comes
from the fact that this other cone is manifestly convex, hence so is the
extended K\"ahler cone{\foot{We would like to thank D.~Morrison for
pointing this out to us.}}.  We will only give a brief sketch of the
argument here and simply use the result in the remainder of the paper. For
a detailed proof, one may consult the mathematics literature
\ref\Kawamata{Y.~Kawamata, ``Crepant blowing-up of 3-dimensional canonical
singularities and its application to degenerations of surfaces,'' Annals of
Mathematics, {\bf 127} (1988), 93--163.}
(see also \Morrison).

There is a one to one correspondence between real cohomology classes of
type (1,1) and line bundles. Namely, given such a class $[\omega]$, one may
find a line bundle $L_{[\omega]}$ such that its first Chern class is
$[\omega]$, and conversely. In order to apply some standard constructions
in algebraic geometry, we will assume that $[\omega]$ is a rational class,
that is it is a class in the intersection of $H^{1,1}(X,{\bf C})$ and
$H^2(X,{\bf Q})$.

With appropriate restrictions, some high tensor power of $L_{[\omega]}$
will have sufficiently many holomorphic sections to define a `good' map to
some projective space, as follows: choose a basis of holomorphic sections
${s_0, s_1, \ldots, s_n}$. Then we get a map $f_{[\omega]}$ from $X$ to
${\bf P}^n$ by sending a point $p$ in $X$ to the equivalence class
$[s_0(p), s_1(p), \ldots , s_n(p)]$. Let us call the image $Y$.  Some
points $p$ may be a common zero for all the $s_i$'s. The collection of such
points is called the {\it base locus} of $L_{[\omega]}$.  Under
$f_{[\omega]}$ the base locus is mapped to the origin in ${\bf C}^{n + 1}$,
so when one projectivises ${\bf C}^{n + 1}$ cycles in the base locus may
get contracted and points may get smeared out. Away from the base locus the
map $f_{[\omega]}$ is an isomorphism.

In order to insure that the image will be a Calabi-Yau that is related to
$X$ by flop transitions at most, we require that $[\omega]$ is {\it
movable}, which means that the base locus is of complex codimension at
least two in $X$.
This condition means that the map $f_{[\omega]}$ is an isomorphism in
codimension one, i.e. the most that can happen is that some two-cycles
contract or some points expand to two-cycles. In particular, the canonical
class of $Y$ must be trivial, so $Y$ is a Calabi-Yau.  Finally, by
construction the holomorphic sections of (some multiple of) $L_{[\omega]}$
get transformed into hyperplane sections, which are the sections of the
line bundle corresponding to the K\"ahler class on $Y$. So the pull-back of
the K\"ahler class on $Y$ is precisely (some multiple of) $[\omega]$.  It
is well known that when $[\omega]$ is taken to be in the original K\"ahler
cone of $X$, $f_{[\omega]}$ will give a smooth embedding of $X$ in
projective space.

Hopefully we have made it plausible that for any rational class $[\omega]$
of type (1,1) on $X$, provided it is movable, we can find a Calabi-Yau $Y$
which has K\"ahler class (a multiple of) $[\omega]$ and is related to $X$
by flop transitions. Conversely, given a Calabi-Yau $Y$ with rational
K\"ahler class $[\varpi]$ and related to $X$ by flops, it can been shown
that the transform of $[\varpi]$ on $X$ is movable. Therefore the extended
rational K\"ahler cone is precisely the rational cone generated by movable
classes of type (1,1). We want to discuss why the property of movability is
preserved under positive linear combinations.

To see this, take two classes $[\omega_1]$ and $[\omega_2]$ and consider
the sum $[\omega] = m [\omega_1] + n [\omega_2]$ where $m$ and $n$ are
positive rational numbers. Since we may multiply $[\omega]$ by any integer,
we may assume that $m$ and $n$ are themselves integer and moreover large
enough for the following argument to apply.  The corresponding line bundle
is
$$ L_{[\omega]} = {L_{[\omega_1]}}^{\otimes m}
\otimes { L_{[\omega_2]}}^{\otimes n} . $$
Take a basis of holomorphic sections $s_i$, $i = 0 \ldots n_1$, for
${L_{[\omega_1]}}^{\otimes m}$. This defines a map to ${\bf P}^{n_1}$.
Similarly choose basis $t_j$, $j = 0 \ldots n_2$, for
${L_{[\omega_2]}}^{\otimes n}$. Then $L_{[\omega]}$ defines a map to ${\bf
P}^{n_1 n_2 + n_1 + n_2}$ by means of the sections $s_i \otimes t_j$. Thus
the base locus of $L_{[\omega]}$ is the union of the base loci of
${L_{[\omega_2]}}^{\otimes m}$ and ${L_{[\omega_2]}}^{\otimes n}$, and in
particular is of codimension at least two.

So we conclude that the cone generated by movable classes is convex, and
therefore that the extended K\"ahler cone is convex.

\subsec{Uniqueness in the extended cone}

Armed with the knowledge that the extended K\"ahler cone is convex, we may
try to employ the argument that was used successfully in section two.  We
start with the local minimum of the central charge at the attractor point
$k_0$. Then on straight lines emanating from $k_0$ the critical points of
$Z$ are all minima as long as they are {\it inside} some K\"ahler
cone. However, a critical point may lie on the boundary between two cones
in which case our argument that it must be a minimum doesn't apply. Recall
that we needed the form $B_{ij}$ to be positive in the direction of the
line. This followed from the physical requirement of the positivity of the
metric inside the moduli space, so that $B_{ij}$ is positive-definite for
all directions tangent to the moduli space. But continuity of the metric
alone does not prevent it from acquiring a zero eigenvalue on the flopping
wall and so in principle $B_{ij}$ may become degenerate there. We are not
aware of a proof of nondegeneracy of the metric (or a counterexample). Thus
it may be possible for a critical point of the central charge along a
straight line to have vanishing second derivative when it lies on the wall
between two K\"ahler cones. Depending on the details of the behaviour of
$Z'$ such a critical point may be a local minimum, maximum or an inflection
point.

By the same token, it may even be that an attractor point $k_0$ is not a
local minimum of the central charge on the extended moduli space, but
rather a saddle point. We therefore switch to a more direct argument.  We
will first consider the case when $k_0$ itself is not on the wall and
consequently it is a local minimum.

Let us again examine the central charge function along a straight line
\straight\ between an attractor point $k_0$ and some other point $k_1$ inside
the extended moduli space. If $k_1$ were another attractor point, then it
would also be a critical point of $Z(t)$ at $t=1$, which is why we will be
looking at the critical points of $Z(t)$. As we know, the only source of
trouble are the points where our line crosses walls of the K\"ahler
cone. Let us first consider the case where only a single wall is crossed
between $t = 0$ and $t = 1$. Suppose that the intersection is at $t =
t_f$. Rather than looking at the central charge $Z(t)$ itself, we will
examine its derivative.  The derivative of $Z$ was
\eqn\firstder{ Z'(t) = (k^3)^{-4/3}
\left( (Q \cdot \Delta k)(k^3)
                  - (Q \cdot k)(\Delta k \cdot k \cdot k)\right).}
The cubic terms in the second term of $Z'$ cancel, so we put
$$Z'(t) = (k^3)^{-4/3}R(t) $$
where $R(t)$ is a polynomial of degree two. As $(k^3)^{-4/3}$ is
nonnegative, we will focus on $R(t)$. By assumption $t = 0$ is a local
minimum for $Z$, i.e. it is a root for $R(t)$ where $R' > 0$.

Recall that at the attractor point we start with a positive value of the
central charge. It is important that for positive $Z$ the only physical
root of $R(t)$ is the one where the first derivative is positive or
possibly zero if such a root is on the flopping wall. The other root of
$R(t)$ is not physical, i.e.\ it lies outside the original K\"ahler
cone. Thus $Z(t)$ has only one critical point in a cone, which is another
proof of uniqueness for a single cone. Now we will see what happens in the
adjacent cone.

First we need to know the point where $Z(t) \sim Q \cdot k(t)$ vanishes,
call it $t_0$. With this definition we may write
\eqn\tzero{ Q \cdot k(t) = Q \cdot k_0 + t \, Q \cdot \Delta k =
     Q \cdot k_0 \, ( 1 -  t / t_0  ) }

When we cross the flopping wall at $t = t_f$ the prepotential changes as
\change
$$ k^3 \to k^3 + c( t - t_f)^3 $$
where $c>0$. So we must also modify $Z'(t)$ for $t > t_f$:
$$ \eqalign{ Z'(t)\big|_{t>t_f}  &= (k^3 + c( t - t_f)^3 )^{-4/3}
    \left( R(t) + c(Q \cdot \Delta k -  Q \cdot k_0)
                                    ( t - t_f)^2 \right) \cr
    &=(k^3 + c( t - t_f)^3 )^{-4/3}
    \left( R(t) + c\ Q \cdot k_0 \left( {t_f \over t_0} - 1 \!
                                   \right)( t - t_f)^2 \right) \cr
    &=(k^3 + c( t - t_f)^3 )^{-4/3} P(t) .} $$

We are interested in the physical roots of $P(t)$ for $t \ge t_f$. As long
as $Z(t)$ is positive those are the roots where $P' \ge 0$. $Z(t)$ is
positive for all $t>0$ if $t_0 < 0 $ while if $t_0 > 0$ it is only positive
for $t < t_0$.  In both cases the constant $ A \equiv c\ Q \cdot k_0 \left(
{t_f \over t_0} - 1 \! \right) $ is negative. First we show that for $A<0$
there are no physical roots of $P(t)$ for $t \ge t_f$.  For that we simply
find the root $t_r$ where $P' \ge 0$ and show that $t_r < t_f$. It will
also imply that $Z(t)$ may not start decreasing and therefore cannot become
negative inside the extended moduli space.

We write $R(t) = a t^2 +b t$ where $b>0$ since $t=0$ is a minimum of $Z$.
Then we need to find roots of the quadratic equation
$$
 P(t) = a t^2 +b t + A (t-t_f)^2 = 0
$$
The root where the derivative $P'(t)$ is nonnegative, if it exists, is
always given by
$$ t_r = { 2 A t_f - b + \sqrt{b^2+4 t_f(-A)(b+a t_f)}\over 2(a+A)} $$
Then there are four cases to consider depending on the value of the
coefficient $a$.

\noindent \item{1.}
$a \leq -b/t_f < 0 $, note that $a + A <0$ and $b + a t_f < 0$, see
Fig.\thinspace 1. In this case there may be no roots, but if there are, we
have
$$
t_r <   { 2 A t_f  - b \over 2(a+A)} \leq { 2 A t_f  + a t_f \over 2(a+A)}
< t_f.
$$
This case is not physical, because the second root of $R(t)$ is inside the
original K\"ahler cone which gives unphysical maximum of the central
charge. We include this case for future reference.

\medskip
\centerline{ \epsfbox{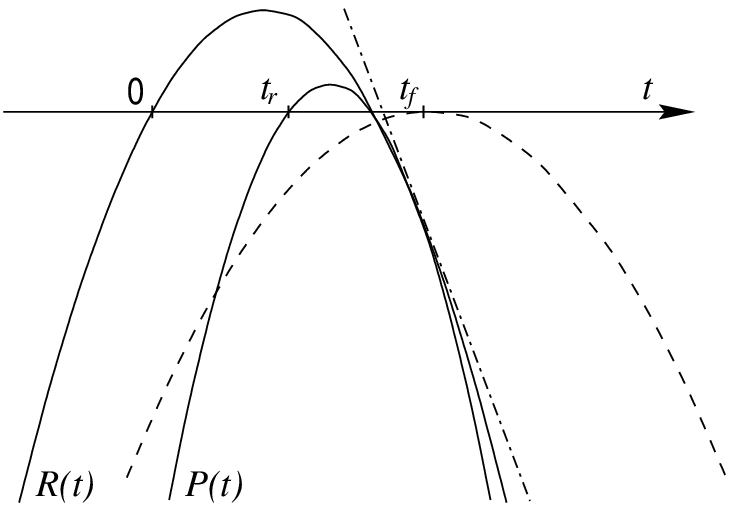}} \nobreak
\centerline{Fig.\ 1}

\item{2.}
$-b/t_f < a < -A$, still $a + A < 0$ but $b + a t_f > 0$. Now there are
always roots and we have
\eqn\ineq{\eqalign{ t_r &<  { 2 A t_f - b +
                      \sqrt{b^2+4 t_f\, a \, (b+a t_f)} \over 2(a+A)} \cr
             &= { 2 A t_f - b + | b + 2  a t_f | \over 2(a+A)} \cr
             & \le { 2 A t_f - b +  b + 2  a t_f  \over 2(a+A)} \cr
             &= t_f.  } }

\item{3.}
$a = -A > 0$. In this case $P(t)$ is linear. It has only one root, which
satisfies
$$
t_r = {a\, t_f^2 \over b + 2 a t_f} <  t_f.
$$

\item{4.}
$-A < a$, see Fig.\thinspace 2. As in item $2$ we replace $-A$ by $a$
inside the square root. The resulting equations are exactly the same as in
\ineq.

\medskip
\centerline{ \epsfbox{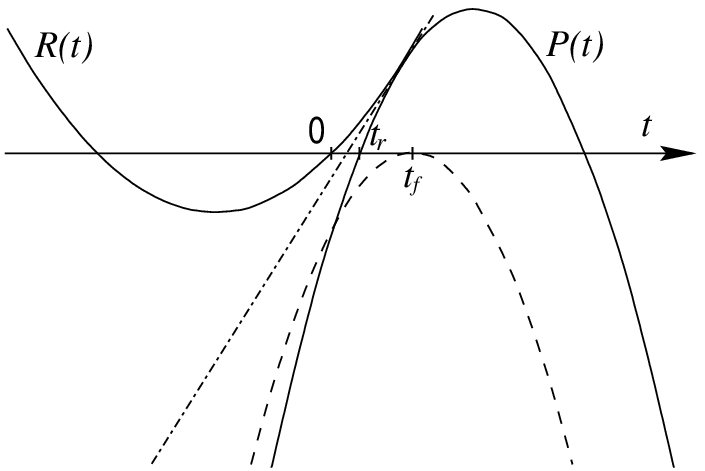}} \nobreak
\centerline{Fig.\ 2.}

What we have found is that $P(t)$ doesn't have physical roots in the new
K\"ahler cone. This means that $Z(t)$ doesn't have critical points there
and therefore continues to grow as we go away from the attractor. It
follows that we need not consider the case when the central charge is
negative.

Crossing several walls can now be handled by induction. The polynomial
$P(t)$ after the wall plays the role of $R(t)$ for the next crossing. We
have shown that $P(t)$ has a root where $P'>0$, but it lies to the left of
the new wall, therefore we are in the same situation as we started.

Let us comment on the (im)possibility of the critical point of $Z(t)$ which
lies on the flopping wall and is a local maximum.  It would correspond to
the situation where $R$ and $R'$ both vanish (recall that $R'$ cannot be
negative at the critical point by continuity and positivity of the metric
away from the flopping wall). Such a point can be described by a situation
in item 4 above with $b=0$, $t_f = 0$ and both roots of $R(t)$ at $0$. As
$R(t)$ is positive to the left of the wall and $P(t)$ is negative to the
right of it, this critical point is a local maximum. However, $R(t)$ cannot
have a double zero at $t_f$ because we have shown that it always has one
root strictly to the left of the wall. Therefore such critical points do
not arise.

Finally, we are left to consider the case when $k_0$ itself lies on the
flopping wall. Then $P(t)$ may be either positive or negative to the right
of the wall. In the former case $Z(t)$ starts growing for $t>0$ and the
analysis we have made earlier carries over with no changes. In the latter
case $k_0$ may be a local maximum in some directions precisely as described
in the previous paragraph. The difference is that we now begin in this
situation and cannot argue that $R(t)$ does not have a double root on the
wall.

In such a case the central charge decreases from $t=0$ and it may
eventually become negative. While it is still positive the first wall
crossing is described essentially by the situation in item 1 above, with
$b=0$. It is then clear that $P(t)$ will have no roots after all subsequent
wall crossings while $Z(t)>0$. Moreover, in every cone $P(t)$ will be a
downward-pointing parabola with the apex to the left of the left wall.

After the central charge becomes negative the discussion changes in two
ways. First, when crossing the wall the constant $A$ in the change from
$R(t)$ to $P(t)$ becomes positive:
$$
R(t) \to P(t) = R(t) + A(t-t_f)^2, \quad A>0.
$$
And second, the physical critical points are now the roots of $R(t)$ where
the first derivative is negative or possibly zero if the root is on the
flopping wall.

Now, $Z(t)$ is decreasing when it becomes negative, therefore the first
critical point after that may only be on the wall of some cone such that
$R(t)$ has a double zero. If we assume that this critical point exists,
right before it $R(t)$ would be a downward pointing parabola with a double
zero on the wall. We can move backwards from it and reconstruct the
polynomials $R(t)$ and $P(t)$ in all the preceding cones. Taking into
account that the constant $A$ is now positive but has to be subtracted, we
see that while $Z$ is negative $R(t)$ in every cone is a downward pointing
parabola with no roots and apex to the right of the right wall. In the cone
where $Z$ crosses zero we obtain a conradiction with the previous analysis
where we have found that the apex of the parabola should be to the left of
the left wall. Therefore, there cannot be multiple attractor points even
when they lie on the walls between K\"ahler cones.

\newsec{Discussion}

In this paper we have demonstrated that a critical point of the central
charge $Z$ is unique if it exists. Moreover, if $Z$ has a minimum (maximum)
at the critical point then it will grow (decrease) along straight lines
emanating from the critical point.  In this section we will discuss two
implications of our result.

If one restricts the moduli to lie inside a single K\"ahler cone then
uniqueness is not surprising. The reason for this derives from the
microscopic interpretation of entropy: it should be possible to reproduce
the entropy of a BPS black hole by a microscopic count of degenerate BPS
states. For the Calabi-Yau black holes considered in this paper, we would
have to count the degeneracy\foot{ 
See \Vafa\ for a discussion of the correct quantity to consider.}
of holomorphic curves within the class specified by the charge vector.  As
we have seen in section two, the macroscopic entropy predicted by the
attractor mechanism is $S = {\pi^2 \over 12} \, Z_0^{3/2}$.  Thus one
expects that the degeneracy of BPS states for large charges, when
supergravity should give a good description, asymptotically approaches
$e^{{\pi \over 24} Z_0^{3/2}}$. In \Vafa\ the count was done for the
special case of elliptic threefolds. The attractor equation in that case
could be solved explicitly and the attractor point was therefore unique (at
least in a single K\"ahler cone).  But even for a general Calabi-Yau one
should not have expected multiple attractor points to occur in a single
K\"ahler cone. Supergravity is well-behaved when the moduli vary only over
a single cone and so the existence of two black hole solutions with
different entropy should have its origin in the possibility of counting
different BPS state degeneracies. But the number of holomorphic curves does
not change inside a K\"ahler cone, so neither should the number of BPS
states. So at least inside a single K\"ahler cone, it is clear that the
entropy should be completely fixed by specification of the charges of the
black hole.

In the extended moduli space however this is not so clear: the number of
curves does change as one crosses a flopping wall. One could therefore
interpret the walls of a K\"ahler cone as a hypersurface of marginal
stability, analogous to the curve of marginal stability in Seiberg-Witten
theory. The puzzle is this: suppose one starts with asymptotic moduli at
some point very far away from the attractor point. Then somehow the
attractor point seems to be aware of the degeneracy of BPS states for the
Calabi-Yau associated with the asymptotic moduli. But when we choose the
asymptotic moduli in a cone that is different from the cone where the
attractor point lies, the degeneracies in the two cones will in general not
be the same, so there is no a priori reason for the absence of multiple
critical points. In the light of our result, one possibility is that the
number of curves changes only very mildly across a transition, mild enough
so that the asymptotic degeneracy in the limit of large charges is not
affected. It would be interesting to check this mathematically.

The existence of multiple attractor points was also thought to be desirable
for the construction of domain walls in five dimensional $N = 2$ gauged
supergravity, along the lines of
\ref\domain{ K.~Behrndt and M.~Cvetic, ``Supersymmetric domain wall world
from D = 5 simple gauged supergravity,'' hep-th/9909058.}.
In that setup the goal is not to minimise the central charge, but to find
extrema of the scalar potential of gauged supergravity, which is
$$ V = - 6 ( W^2 - {3\over 4} g^{ij} \partial_i W \partial_j W). $$
In the above, $W = Q_i k^i$ where $k^i$ are the usual K\"ahler moduli and
$Q_i$ are the gravitino and gaugino charges under the $U(1)$ that is being
gauged. Even though the interpretation is different, $W$ is numerically the
same as what we have called $Z$ before and the supersymmetric critical
points of $V$ are also critical points of $W$ \FiveDims. At a
critical point $W_0$ of $W$ the five dimensional supergravity solution is
anti-De Sitter space with cosmological constant equal to $-6 W_0^2$. To
construct a domain wall, one would like to have two critical points $k_0$
and $k_1$ of $W$. Then one could write down a supergravity solution that
interpolates between two different anti-De Sitter vacua, with the
asymptotic values of the moduli being $k_0$ on one side of the wall and
$k_1$ on the other. It was hoped that this might lead to a supergravity
realisation of the Randall-Sundrum scenario
\ref\RS{ L.~Randall and R.~Sundrum, ``An alternative to compactification,''
hep-th/9906064.}.
Unfortunately as we have seen, this construction does not appear to be
possible, at least in its simplest form, because of the absence of multiple
(supersymmetric) critical points.

Finally, the attractor mechanism in four dimensions is somewhat similar to
the five dimensional mechanism considered in this paper. It would be
interesting if our methods could be used to shed some light on this
important problem as well.

\bigskip

{\bf Acknowledgements:} We would like to thank M.~Bershadsky, R.~Gopakumar,
S.~Nam, A.~Klemm, T.~Pantev, and especially D.~Morrison and C.~Vafa for
valuable discussions and support. The work of SZ was supported in part by
the DOE grant DE-FG02-91ER40654.

\listrefs

\bye